\documentclass[%
 preprint,
 amsmath,amssymb,
 aps,
prb,
]{revtex4-1}
\usepackage{graphicx}% Include figure files
\usepackage{dcolumn}% Align table columns on decimal point
\usepackage{bm}% bold math
\usepackage{cleveref }
\usepackage{multirow}
\begin{document}

\title{THz generation using a reflective stair-step echelon}% Force line breaks with \\

\author{Benjamin K. Ofori-Okai$^{1\ddag}$, Prasahnt Sivarajah$^{1\ddag}$, W. Ronny Huang$^2$,}
\author{Keith A. Nelson$^1$}%
 \email{Corresponding author; e-mail: kanelson@mit.edu}
\affiliation{$^1$Department of Chemistry, Massachusetts Institute of Technology, Cambridge, M.A 02139, USA\\
$^2$Department of Electrical Engineering and Computer Science and Research Laboratory of Electronics, Massachusetts Institute of Technology, Casmbridge, M.A 02139, USA\\
$^{\ddag}$ These authors contributed equally to this work}


%\affiliation{
%affiliation{
%\affiliation{%


\date{\today}% It is always \today, today,

\begin{abstract}
We present a novel method for THz generation in lithium niobate using a reflective stair-step echelon structure. The echelon produces a discretely tilted pulse front with less angular dispersion compared to a high groove-density grating. The THz output was characterized using both a 1-lens and 3-lens imaging system to set the tilt angle at room and cryogenic temperatures. Using broadband 800 nm pulses with a pulse energy of 0.95 mJ and a pulse duration of 70 fs (24 nm FWHM bandwidth, 39 fs transform limited width), we produced THz pulses with field strengths as high as 500 kV/cm and pulse energies as high as 3.1 $\mu$J. The highest conversion efficiency we obtained was 0.33\%. In addition, we find that the echelon is easily implemented into an experimental setup for quick alignment and optimization.
%\begin{description}
%\end{description}
\end{abstract}

\maketitle

\section{Introduction}
The generation of high power THz radiation has led to the development of THz nonlinear optics and spectroscopy \cite{Hoffmann2011, Tanaka2011, Hwang2013}. Nonlinear THz-driven responses in the 0.1-5 THz range have been observed in semiconductors \cite{Hoffmann2009, Hebling2010, Hirori2011a, Fan2013}, graphene \cite{Hwang2013}, quantum confined systems \cite{Hirori2010, Shinokita2010}, liquids \cite{Hoffmann2009a}, gases \cite{Fleischer2011, Fleischer2012}, molecular crystals \cite{Jewariya2010}, and correlated electron materials \cite{Grady2013, Liu2012}. This advancement has been primarily enabled by the ability to use lithium niobate (LN) for THz generation by optical rectification. Prior to this, the most common sources of high-power THz radiation were either free-electron lasers \cite{vanderMeer2004} which have limited accessibility, QCLs \cite{Williams2007} which have low peak fields, and gas lasers which have long pulse durations. Compared to other widely used nonlinear materials like zinc telluride (ZnTe) or gallium phosphide (GaP), LN has a larger nonlinear coefficient ($d_{ZnTe} \approx 67\textrm{ pm/V, }d_{GaP} \approx 44 \textrm{ pm/V, }d_{LN} \approx 198$ pm/V \cite{Hebling2008}) for optical rectification, which leads to greater THz generation efficiency. The major drawback of LN, however, is the large index mismatch between the ultrafast infrared (IR) or near infrared (NIR) pump pulse that is rectified and the THz field that is produced ($n_{LN}^{gr} \sim 2.2$, $n_{LN}^{THz} \sim 5$). This causes the pump pulse to outrun the generated wave, and the THz field is consequently emitted as a cone of Cherenkov radiation \cite{Auston1984}. While this prevents collinear phase matching in LN, it is possible to efficiently generate THz fields using noncollinear phase matching with tilted optical pulse fronts \cite{Hebling2002, Stepanov2003}. Typically, this is accomplished by using a grating to introduce angular dispersion into the beam, resulting in a continuously tilted pulse front \cite{Hebling1996}. This tilted pulse front is them imaged \cite{Kunitski2013} into the LN prism to produce a pulse with the correct tilt angle so the optical intensity front moves laterally in the LN at a speed that allows matching of the THz phase velocity.

To date, there have been considerable experimental \cite{Yeh2007, Stepanov2008, Nagai2009, Bodrov2013, Huang2013, Fulop2014, Huang2014, Wu2014} and theoretical \cite{Fulop2010, Fulop2011, Ravi2014} efforts aimed at improving the generation efficiency of the tilted pulse front technique. Numerical studies suggest \cite{Fulop2010, Fulop2011} that the three major factors limiting the generation length in the LN crystal come from (1) the degradation of the pulse duration away from the image plane due to the angular dispersion introduced \cite{Martinez1986} and (2) the imperfect imaging of the tilted pulse front from the grating. In addition to these, recent work \cite{Ravi2014, Bodrov2013, Lombosi2015} has proposed that (3) the inherent coupling between the generated THz radiation and the pump pulse leads to detrimental effects which significantly reduce the interaction length of the pump pulse and necessarily limit the generation efficiency. 

While the coupling effects are difficult to manage practically, several suggestions have been proposed to mitigate the other issues. The first problem can be compensated for by using pulses with smaller bandwidths and hence longer transform limited pulse durations, but this reduces the generated THz bandwidth since there are fewer frequency components to use for optical rectification. Moreover, many high-field THz generation systems are based on amplified Ti:sapphire lasers whose pulse durations are tens of femtoseconds. Such short pulses are often desirable in a nonlinear THz spectroscopy measurement or for electro-optic sampling. For these systems, increasing the transform limited duration by filtering would come at the cost of both time resolution and reduced optical pulse energy. It has also been proposed \cite{Palfalvi2008} that using a direct-contact grating would mitigate the problem of imperfect imaging. However, this approach reduces the effective generation length compared to that reached by imaging of the beam from a typical grating. This is because when imaging the grating, temporal focusing of the pump pulse and associated optimization of THz generation efficiency occurs on both sides of the image plane over a total distance $2L_{d}$ where $L_{d}$ is the dispersion length \cite{Diels2006}. However, with the contact grating method, the length over which the optical pulse duration is short is reduced by a factor of 2. Furthermore, this still requires the use of a highly dispersive grating that leads to limited generation lengths, especially when using pulses with considerable bandwidth.

\begin{figure}[ht!]
\centerline{\includegraphics[width=5.25in]{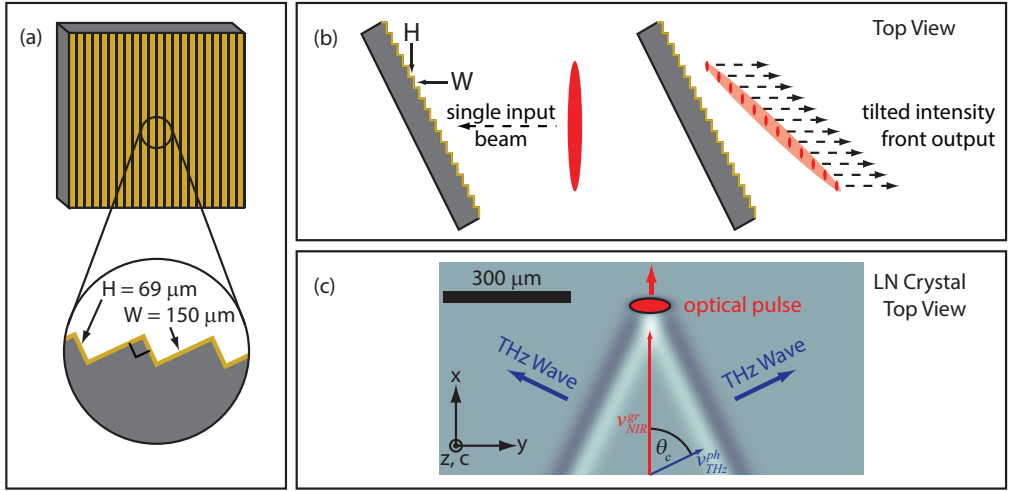}}
\caption{(a) Schematic illustration of the stair-step echelon used for THz generation. The overall dimensions of the echelon are 50 mm $\times$ 50 mm $\times$ 10 mm. The height and width of each echelon step are $H = 69$ $\mu$m and $W = 150$ $\mu$m, respectively. (b) Method of operation: The echelon is set so that the width (``horizontal") direction of the steps is at normal incidence to the incident beam. Each step produces a single reflected ``beamlet" whose pulse is time delayed relative to an adjacent beamlet pulse due to the additional $2H$ path length. (c) Calculated Cherenkov wavefronts produced by a single beamlet pulse propagating through a LN crystal. The beamlet is polarized along the extraordinary $(z)$ axis, which is parallel to the LN crystallographic $c$-axis. The spatially and temporally shifted beamlets all generate THz wavefronts that superpose constructively on one side.
\label{ech_schematic}}
\end{figure}

An alternate scheme which ideally avoids the introduction of large amounts of angular dispersion is a discretely tilted pulse front from a reflective echelon \cite{Minami2013}. The reflective echelon consists of many small steps that act as mirrors and split a single input beam into many smaller time-delayed ``beamlets", which together create a discretely tilted pulse front as shown in figure \ref{ech_schematic}(a)-(b). The beamlets are then imaged into the LN prism and act as near-line sources to produce Cherenkov sheets of THz radiation. As the spacing of the imaged steps is small compared to THz wavelengths, the discretely tilted pulse front is effectively continuous over a portion of the THz bandwidth. Here, each beamlet is assumed to generate a THz field independently, provided that the duration of each beamlet is short relative to the time delay between them \cite{Stepanov2005a}. Figure \ref{ech_schematic}(c) shows results of a simple 2D calculation of the THz sheet generated from a single infinitely tall beamlet with a Gaussian spatiotemporal profile \cite{Feurer2007}. In this case, the frequency content of the resulting THz field is set by the beamlet's lateral size. When the tilt angle is set appropriately, the THz fields resulting from the many responses superpose constructively on one side, leading to coherent field addition and efficient THz generation.

In conventional THz generation with a highly dispersive grating, a 1-lens imaging system is used to image the pump beam into the LN prism, and beam divergence is largely ignored because angular dispersion and pulse front distortions in the image plane are more far detrimental \cite{Fulop2010}. For the echelon, the beamlet spot size $w_{0}$ in the image plane is large relative to the pump pulse wavelength $\lambda$ and the Rayleigh range $z_{R}=\pi w_{0}^2/\lambda$ is independent of the pulse duration. As a result, the beamlets can be of shorter pulse duration and can propagate for a long distance $2z_{R}$ without appreciable divergence. Pump beam divergence and THz absorption losses in the LN prism then become limiting factors. This suggests additional benefits by using multiple lenses for imaging to produce collimated beamlets at the image plane that allow exploitation of the long Rayleigh range, and by cryogenic cooling of the LN prism \cite{Fulop2011, Blanchard2014} to reduce THz absorption which is due primarily to phonon damping.
	
In this work, we demonstrate THz generation using a discretely tilted pulse front produced by a reflective echelon as described above. Experiments were performed using a 1-lens imaging system and a 3-element varifocal zoom-lens capable of continuously tunable demagnification from $M$ = 0.16 to $M$ = 0.29 \cite{Back1954} at both room temperature (RT) and cryogenic temperatures (CT, $T \sim$ 100 K). Using discrete optical pump beams to deliver spatially and temporally shifted pulses for THz generation in LN or lithium tantalate has been reported based on different approaches including beamsplitting to produce just two beams \cite{Koehl2001}, spatiotemporal pulse shaping \cite{Koehl2001a, Feurer2003}, and transmission through a homebuilt stairstep echelon \cite{Yeh2007a}. In these cases the THz wave remained inside the generation crystal and the conversion efficiency was not measured. Our measurements of the output THz power and field strength show that by using the echelon, it is possible to achieve conversion efficiencies better than those obtained using a grating at room temperature using broadband pulses with sub 100 fs duration from a commercial Ti:sapphire laser. It may also be possible to use pulses with even shorter durations ($<$50 fs) since the discretely tilted pulse front should not suffer from the adverse effects introduced by angular dispersion. 

\section{Experimental Setup}

A schematic illustration of the echelon is shown in figure \ref{ech_schematic}. The echelon was fabricated by Sodick F.T. (Japan) by cutting a stair-step pattern into a block of steel followed by gold plating for high reflectivity. The step width and height were $W = 150.0 \pm 0.2$ $\mu$m and $H = 69.00 \pm 0.02$ $\mu$m, respectively. As shown in figure \ref{ech_schematic}(a), the height determines the time delay between adjacent pulses which is given by $t = 2H/c = 460.31 \pm 0.07$ fs, and the width determines the initial width of the beamlet (whose size in the orthogonal direction is $\sim$ 9 mm) that is reflected by each step surface. The echelon was rotated such that the direction along the 150-$\mu$m widths of the steps was normal to the incident beam path. Upon reflection, the input beam was split into many beamlets, resulting in a discrete tilted pulse front with a tilt angle $\theta = \textrm{arctan}(2H/W) = 42.6^{\circ}$. The tilt angle was then adjusted by demagnification while accounting for the change in tilt angle upon transmission into the LN prism \cite{Hebling2008}. For a given demagnification factor M, the final tilt angle, $\gamma$, is given by

\begin{equation}
\gamma = \textrm{arctan}\left(\frac{2HM}{Wn_{LN}^{gr}} \right)
\label{Ech_tilt_angle}
\end{equation}

\noindent where $n_{LN}^{gr}$ is the group index of the NIR pulse in LN.

\begin{figure}[htbp]
\centerline{\includegraphics[width=5.25in]{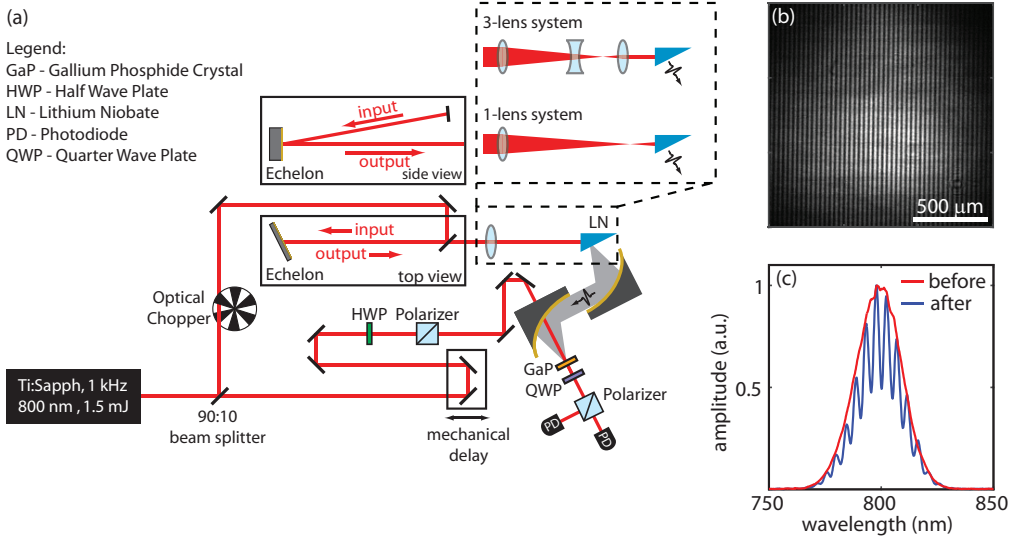}}
\caption{(a) Experimental setup. The output of an amplified Ti:Sapphire laser system was split 90:10 for the THz generation arm and EO sampling arms, respectively. The portion used for THz generation was directed through an optical chopper and toward the echelon which was imaged with a demagnification of $\sim$5 into the LN prism. The THz output from the LN prism was collected using a pair of off-axis parabolic mirrors and focused into a 200-$\mu$m active, 2-mm inert GaP crystal pair for EO sampling. (b) Image of echelon recorded with a camera placed at the image plane. The bright vertical strips are the individual echelon steps. (c) Spectra of the pump beam before (red) and after (blue) reflecting off the echelon. The strong modulation indicates that the single beam has been split into time delayed coherent pulses.}
\label{ech_gen_setup}
\end{figure}

The setup used in the experiments is represented in figure \ref{ech_gen_setup} (a). All measurements were performed using NIR pulses from an amplified Ti:Sapphire laser system (800 nm center wavelength, 1.5 mJ pulse energy, 1 kHz repetition rate, 24 nm FWHM bandwidth, 70 fs pulse duration). The beam was split 90:10 for THz generation and electro-optic (EO) sampling. The generation beam was chopped at 500 Hz using an optical chopper and tilted toward the echelon as shown in the side view inset of the figure. The reflected beamlets were imaged into a right-angle LN prism with a demagnification factor of $M\approx 5$ using either an $f = 8$ cm achromatic lens or 3-element zoom-lens imaging system with $f_{1} = 30$ cm, $f_{2} = -7.5$ cm, and $f_{3} = 7.5$ cm. The 1/$e^{2}$ radius of the optical pump beam at the face of the LN prism was approximately 1.0 mm. As shown in \ref{ech_gen_setup}(b), the incident light at the LN prism consists of $\sim 60$ individual beamlets, each one 30 $\mu$m wide. The effect of the temporal delay of the beamlets is seen in \ref{ech_gen_setup}(c), which shows the spectrum of the pump light before and after reflection. The initially smooth spectrum develops a strong $4.4 \pm 0.2$ nm spectral modulation after reflection as is expected from pulses delayed by 460 fs. The spectra were obtained by directing diffuse scattering of the light into a spectrograph. The spectrum of the light reflected from multiple echelon steps provides an accurate measure of the echelon step height, but the spectrum of any individual beamlet is that of the original pump beam, without spectral modulation.
	
The resulting THz output was collected using a 3-inch diameter, 3-inch effective focal length (EFL), 90$^{\circ}$ off-axis parabolic reflector, and focused using a 3-inch diameter, 2-inch EFL parabolic reflector into a GaP crystal. The EO sampling optical beam was variably delayed, attenuated using a half-wave plate and polarizer, directed through a hole in the second THz parabolic reflector, and overlapped with the THz pulse in the GaP crystal. The transmitted EO sampling beam was then passed through a quarter wave plate and polarizing beam splitter, and the signals were detected using balanced photodiodes and analyzed using a National Instruments Data Acquisition card \cite{Werley2011}.

\section{Results}
\subsection{THz Characterization Procedure}

\begin{figure}[htbp]
\centerline{\includegraphics[width=5.25in]{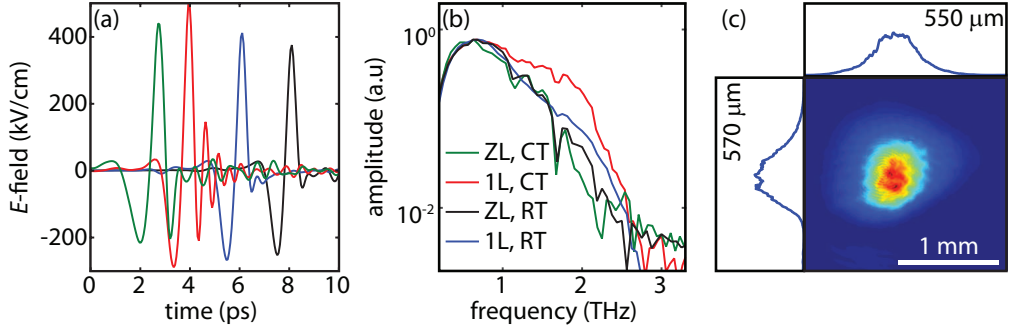}}
\caption{(a) Free-space electro-optic sampling data obtained using a 200 $\mu$m active, 2 mm inert GaP crystal-pair for both the 1-lens (1L) and zoom-lens (ZL) imaging systems at both RT and CT. (a) Time-domain waveforms of THz pulse with the the highest peak THz field around 500 kV/cm using 1L at CT. (b) Normalized magnitudes of the Fourier transform of the THz time-domain waveform. In all cases, the peak frequency is approximately 0.63 THz, the full width-half maximum bandwidth is in the range $\Delta\nu \approx$ 0.8-1.3 THz, and the spectral amplitude falls to 10\% of the peak at $\sim$ 2.3 THz using 1L at CT, and at $\sim$ 1.6 THz for all other cases. (c) Image of the THz intensity profile at the focus of the GaP crystal using the ZL at CT. The blue traces along the top and side are line-outs through the center of the THz spot. These were fit to Gaussian functions and the extracted values represent the $1/e^{2}$ radii along the horizontal and vertical directions.}
\label{THz_timetrace_spectrum_spotsize}
\end{figure}

To begin characterizing the THz output of the system, we measured the THz time-dependent waveform by EO sampling and computed the corresponding Fourier transform, yielding the results shown in figure \ref{THz_timetrace_spectrum_spotsize}(a)-(b). The data shown were obtained using maximum pump power for the 1-lens and zoom-lens at RT and CT, and the results are summarized in table \ref{tab:THz_results} below. 

The highest peak THz electric field ($E$-field) measured was 500 kV/cm when using 1-lens at CT. In all cases, the peak frequency was 0.63 THz and the mean frequencies ranged from 0.9 THz to 1.1 THz.  Our peak frequency is slightly lower than the peak frequencies typically obtained using a continuous tilted pulse front, which is likely a result of the finite size of the beamlets in the LN prism. Assuming independent THz generation by each of the beamlets, an estimate of the peak THz frequency, $\nu_{p}$, can be made (neglecting frequency-dependent THz absorption in the LN crystal) according to \cite{Feurer2007}: 

\begin{equation}
\nu_{p} \approx \frac{c_0}{\pi n_{LN}^{THz} w_{0}}
\label{Peak_THz_Freq}
\end{equation}

\noindent where $n_{LN}^{THz} = 4.96$ is the THz refractive index of LN and $w_{0}$ is the spot size of the beamlet. Using $w_{0} = 30$ $\mu$m, we find $\nu_{p}$ = 0.64 THz in almost exact agreement with our experiment. The full-width-half maximum of the THz spectrum was $\Delta\nu \approx$ 0.8 THz in all cases except for the 1-lens at CT, which had $\Delta\nu \approx$ 1.3 THz. These values are all comparable to values obtained using 800 nm pump light with a grating to produce a continuous tilted pulse front \cite{Hwang2014}. For the echelon, the bandwidth of the resulting THz pulses depends on the available wave vector components of the beamlets, with smaller beamlets having more content. Any additional divergence in the full beam through the image plane will also lead to variations in the tilt angle and thus to changes the THz bandwidth \cite{Hebling2004}.

\begin{table}[ht!]
\caption{Summary of THz generation results} 
\centering
	\begin{tabular}{c c c c c}
	\hline 
			
Setup 
& \begin{tabular}[c]{c}Peak field\\(kV/cm)\end{tabular}
& \begin{tabular}[c]{c}Peak frequency\\(THz)\end{tabular}
& \begin{tabular}[c]{c}Mean frequency\\(THz)\end{tabular}
& FWHM (THz)
 \\ \hline

1L RT
& 410
& 0.63
& 0.94
& 0.83
\\

1L CT
& 500
& 0.63
& 1.10
& 1.33
\\

ZL RT
& 375
& 0.63
& 0.96
& 0.80
\\

ZL CT
& 440
& 0.63
& 0.90
& 0.79
\\
\hline
	\end{tabular}\label{tab:THz_results}
\end{table}

To further characterize the THz output, we measured the spatiotemporal evolution of the THz pulse at the focus of the second parabolic reflector \cite{Bodrov2013}. Instead of focusing the EO sampling beam, it was left collimated and allowed to over-fill the GaP crystal which was imaged using a 1-lens imaging system through a quarter waveplate and Wollaston prism onto a camera resulting in a pair of images. A time series of these images was collected with the THz field present and the THz field absent and then balanced to produce a movie (see Multimedia 1). 

Figure \ref{THz_timetrace_spectrum_spotsize}(c) shows an image of the spot obtained at the peak of the time-domain waveform using the zoom-lens at CT. The extracted signal is proportional to $E(x,y)$ and was squared to obtain an intensity distribution $I(x,y) \propto E_{THz}^{2}(x,y)$ . The traces along the top and the side of the 2D image correspond to vertical and horizontal cross sections through the center of the spot. These were fit to Gaussian functions and values of 0.57 mm and 0.55 mm were obtained for the $1/e^{2}$ radii along the vertical and horizontal dimensions, respectively.

\subsection{THz pulse energy and conversion efficiency}

While the peak E-field and bandwidth of the THz pulse are important for driving nonlinear effects in materials, a more generalized metric for comparison to other THz generation schemes is the pump-to-THz conversion efficiency. In order to accurately determine this, we calculated the THz pulse energy using parameters obtained by EO sampling measurements and the THz intensity profile assuming Gaussian spatial and temporal profiles. The free space pulse energy, $U$, is given by \cite{Reid2005}

\begin{equation}
U = \frac{1}{2gT}\pi c_0 \epsilon_0 n_{GaP}^{THz} E_0^2 w_x w_y \tau_0
\label{Pulse_Energy}
\end{equation}

\noindent where g = 0.94 is a correction factor assuming a Gaussian temporal shape, T = 0.72 is the power transmission coefficient into the GaP crystal, $E_{0}$ is the peak field strength in the EO crystal, $\tau_{0}$ is the intensity full width-half maximum pulse duration, $w_{x}$ and $w_{y}$ are the 1/$e^{2}$ radii along $x$- and $y$-directions, and $n_{GaP}^{THz}$ = 3.24 is the THz refractive index of GaP. 

\begin{figure}[ht!]
\centerline{\includegraphics[width=5.25in]{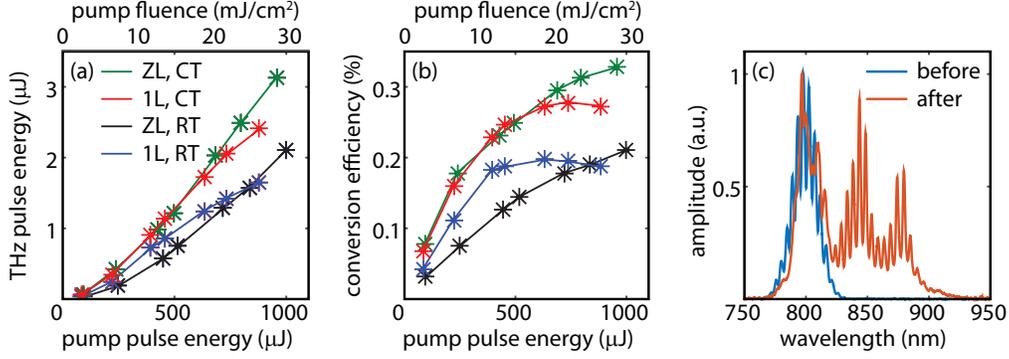}}
\caption{(a) THz pulse energy as a function of NIR pump pulse energy using the 1-lens (1L) and zoom lens (ZL) at both RT and CT, calculated from measured THz field strengths and profiles. For all measurements, the pump 1/$e^{2}$ radii were $w_{x}=w_{y}\sim1.0$ mm entering the LN prism. (b) Calculated THz conversion efficiency. (c) Spectra of the NIR pump pulse of 1mJ reflected from the echelon, both before and after passing through the LN prism. The strongly modulated pump pulse experiences significant spectral red-shifting and broadening, which is indicative of cascading during THz generation in the LN prism.}
\label{pulse_energy_conv_efficiency_spectra}
\end{figure}

By measuring the peak THz field at various pump pulse energies using the 1-lens and zoom-lens imaging systems at both RT and CT, we calculated the THz pulse energy, plotted in figure \ref{pulse_energy_conv_efficiency_spectra}(a), and the corresponding conversion efficiency, plotted in figure \ref{pulse_energy_conv_efficiency_spectra}(b). In all cases, below 500 $\mu$J pump pulse energy (14.4 mJ/cm$^{2}$ fluence) the THz pulse energy increased quadratically with pump pulse energy, corresponding to a linear rise in the conversion efficiency. For the 1-lens system, above 700 $\mu$J pump energy (20.2 mJ/cm$^{2}$ fluence) the THz pulse energy grew linearly with pump pulse energy at both RT and CT, which is also indicated by a saturation of the conversion efficiency. Previous studies using 800 nm pulses observed saturation in the conversion efficiency at lower pumping fluence \cite{Bodrov2013, Wu2014}, which was attributed to unwanted effects from self-phase modulation \cite{Ravi2014} and three-photon absorption \cite{Wu2014}. In contrast, the pulse energy continues super-linear growth when imaged using ZL system, which is confirmed by the continued increase in the conversion efficiency.

The data are summarized in table \ref{tab:THz_conv_eff}. From these values, we observed an enhancement in the conversion efficiency from RT to CT by a factor of 1.4$\times$ for the 1-lens imaging system and 1.6$\times$ for the zoom-lens. We obtained a peak conversion efficiency of 0.33\% using 0.95 mJ energy (27.5 mJ/cm$^{2}$ fluence) pulses using the zoom-lens at CT, and our highest THz pulse energy exceeded 3.1 $\mu$J . This maximum almost exceeds the record 0.35\% using 800-nm center wavelength pulses \cite{Blanchard2014} (based on pyroelectric detector measurement). In that case, the tilted pulse front was produced by a grating, and a much longer transform-limited pulse duration ($>200$ fs) was used. With the echelon, we achieved a comparable conversion efficiency using a much shorter transform-limited pulse duration where the grating performance is expected to be limited \cite{Fulop2010}. The energy of the THz pulses resulting from echelon-based generation was also measured using a TK power meter and a Microtech pyroelectric detector. The values obtained from these other methods both exceeded those calculated using the parameters assuming a Gaussian beam by roughly a factor of 2. The numbers reported here, therefore, may represent a lower bound on the THz pulse energy and conversion efficiency of the echelon generation method. In addition, the conversion efficiencies measured using these instruments followed the trends indicated below, with relative values in good agreement. EO sampling measurements made with GaP or ZnTe crystals of different thickness gave absolute THz field levels and conversion efficiencies also in good agreement with those reported in table \ref{tab:THz_conv_eff}. 

\begin{table}[h!]
\caption{Summary of THz generation results} 
\centering
	\begin{tabular}{c c c c c}
	\hline 
			
Setup 
& THz pulse energy ($\mu$J)
& Conversion Efficiency (\%)
& Cooling Enhancement
 \\ \hline

1L RT
& 1.6
& 0.19
& \multirow{2}{*} {1.4$\times$}
\\

1L CT
& 2.4
& 0.27
\\

ZL RT
& 2.1
& 0.21
& \multirow{2}{*} {1.6$\times$}
\\

ZL CT
& 3.1
& 0.33
\\
\hline
	\end{tabular}\label{tab:THz_conv_eff}
\end{table}

Finally, we characterized the spectrum of the NIR pump pulse used for THz generation. This is plotted in figure \ref{pulse_energy_conv_efficiency_spectra}(c), and shows significant spectral broadening as well as a red-shift in the center wavelength of the modulated spectrum. This red-shift is characteristic of the THz generation process \cite{Yeh2007, Stepanov2007, Nagai2009, Jewariya2009}, which occurs through difference-frequency mixing among the optical spectral components \cite{Cronin-Golomb2004, Stepanov2007, Ravi2014} and which can be cascaded through multiple difference-frequency steps, leading greater than 100\% photon conversion efficiencies.

\subsection{Discussion of Comparison}

In comparing the measured conversion efficiencies, the 1-lens and zoom-lens imaging systems show markedly different behavior with respect to pump pulse energy. At both RT and CT, the 1-lens system conversion efficiency appears to saturate, while the zoom-lens system conversion efficiency increases due to continued superlinear growth. In addition, the 1-lens system shows a noticeable increase in the THz bandwidth at CT as a result of the dramatic reduction in THz absorption at higher frequencies \cite{Fulop2011}, while the bandwidth obtained from the zoom-lens system does not noticeably change. This behavior could be due to the larger divergence and wavevector spread of the beamlets with the 1-lens imaging system, which would allow phase matching over a wider range of frequencies as described in in Sec. 3.1. In contrast, the divergence of the beamlets is diffraction limited with the zoom-lens system, and hence there is negligible change in the generated bandwidth. The beam divergence in the 1-lens system also leads to a range of tilt angles that expands more rapidly in front and back of the image plane. While this may diminish the overall conversion efficiency, it may at the same time increase conversion into higher-frequency THz spectral components. 

Our THz generation system has some similarities to a phased antenna array \cite{Hansen2009}. Treating each beamlet as an element of such an array, the superposition of THz waves along the Cherenkov angle is reduced when $\lambda_{THz} \leq 30$ $\mu$m or $f_{THz} \geq 2$ THz, as energy is diffracted into higher orders by the imposed periodic structure. A smaller echelon step width may therefore improve the conversion efficiency by reducing the amount of diffraction of high frequency waves. This would require a proportional reduction in the step height. The reductions would ultimately be limited by the pulse duration since for pulses with durations $\tau > 2H/c$, it is possible that the echelon would act as an echelle grating that produces a beam with low angular dispersion because of its significantly reduced groove density. In that case, the larger demagnification used here ($M\approx5$) would increase the angular dispersion to be the same as that from a conventional grating. At the image plane the pump pulse from the echelon would therefore suffer the same degradation, limiting the generation efficiency. Our assumptions that the discrete tilted pulse front properties are insensitive to the pump pulse duration and that each beamlet acts as an independent line source in the LN crystal, with a Rayleigh range given by its spot size, would no longer be valid in this limit. 

Careful theoretical treatment of THz generation using a continuous tilted pulse front from a grating has shown that the dramatic modification of the pump spectrum is significantly influenced by the nonlinear coupling between the THz and optical fields which candramatically reduce the effective THz generation length \cite{Ravi2014, Lombosi2015}. Similar calculations are under way for the echelon generation that we have demonstrated here, and the results may help understand the limitations that are faced in this approach. 

\section{Conclusions}

We have demonstrated efficient THz generation using an easily implemented 800-nm discrete tilted pulse front produced by a reflective stair-step echelon structure. We produced THz pulses with field strengths as high as 500 kV/cm and pulse energies exceeding 3 $\mu$J. The peak conversion efficiency at cryogenic temperature was 0.33\%, which is an improvement over values routinely reported in the literature for sub-100 fs pulse durations. Further improvements to the echelon could be made to increase the THz output, such as reducing the step width to facilitate the generation of higher frequencies. 

It is worth noting that higher conversion efficiencies have been observed using even longer transform-limited pulses at 1030 nm center wavelength \cite{Fulop2014, Huang2013, Huang2014}. When generating using 1030 nm, saturation of the conversion efficiency occurs at higher intensity because free carriers are generated by four-photon absorption, rather than three-photon absorption at 800 nm \cite{Hoffmann2007}. Furthermore, the high conversion efficiencies for these longer transform-limited pulses were attributed mainly to the reduced pump pulse bandwidth, which increases the coherence length and decreases the effects from stimulated Raman scattering \cite{Ravi2014, Nagai2012, Fulop2010, Fulop2011}. It is likely that the use of an echelon would allow shorter pump pulse durations to be used at similar pump wavelengths, resulting in high conversion efficiencies with greater THz bandwidths. 

\section{Acknowledgments}

The authors thank Dr. Harold Hwang, Dr. Christopher Werley, and Mr. Koustuban Ravi for fruitful discussions. Funding for this work was provided by the Samsung GRO program. B.K.O. was supported in part by a National Science Foundation Graduate Research Fellowship Program (NSF GRFP), P.S. was supported in part by a Natural Sciences and Engineering Research Council of Canada Postgraduate Scholarship (NSERC PGS D), and W.R.H. was supported in part by a National Defense Science and Engineering Graduate (NDSEG) Fellowship.

\bibliographystyle{osajnl}

\begin{thebibliography}{10}
\newcommand{\enquote}[1]{``#1''}

\bibitem{Hoffmann2011}
M.~C. Hoffmann and J.~A. F\"{u}l\"{o}p, \enquote{{Intense ultrashort terahertz
  pulses: generation and applications},} Journal of Physics D: Applied Physics
  \textbf{44}, 083001 (2011).

\bibitem{Tanaka2011}
K.~Tanaka, H.~Hirori, and M.~Nagai, \enquote{{THz Nonlinear Spectroscopy of
  Solids},} IEEE Transactions on Terahertz Science and Technology \textbf{1},
  301--312 (2011).

\bibitem{Hwang2013}
H.~Y. Hwang, N.~C. Brandt, H.~Farhat, A.~L. Hsu, J.~Kong, and K.~A. Nelson,
  \enquote{{Nonlinear THz conductivity dynamics in P-type CVD-grown graphene.}}
  The Journal of Physical Chemistry B \textbf{117}, 15819 (2013).

\bibitem{Hoffmann2009}
M.~C. Hoffmann, J.~Hebling, H.~Y. Hwang, K.-L. Yeh, and K.~A. Nelson,
  \enquote{{Impact ionization in InSb probed by terahertz pump-terahertz
  probe spectroscopy},} Physical Review B \textbf{79}, 161201 (2009).

\bibitem{Hebling2010}
J.~Hebling, M.~C. Hoffmann, H.~Y. Hwang, K.-L. Yeh, and K.~A. Nelson,
  \enquote{{Observation of nonequilibrium carrier distribution in Ge, Si, and
  GaAs by terahertz pump-terahertz probe measurements},} Physical Review B
  \textbf{81}, 035201 (2010).

\bibitem{Hirori2011a}
H.~Hirori, K.~Shinokita, M.~Shirai, S.~Tani, Y.~Kadoya, and K.~Tanaka,
  \enquote{{Extraordinary carrier multiplication gated by a picosecond electric
  field pulse.}} Nature Communications \textbf{2}, 594 (2011).

\bibitem{Fan2013}
K.~Fan, H.~Y. Hwang, M.~Liu, A.~C. Strikwerda, A.~J. Sternbach, J.~Zhang,
  X.~Zhao, X.~Zhang, K.~A. Nelson, and R.~D. Averitt, \enquote{{Nonlinear
  Terahertz Metamaterials via Field-Enhanced Carrier Dynamics in GaAs},}
  Physical Review Letters \textbf{110}, 217404 (2013).

\bibitem{Hirori2010}
H.~Hirori, M.~Nagai, and K.~Tanaka, \enquote{{Excitonic interactions with
  intense terahertz pulses in ZnSe/ZnMgSSe multiple quantum wells},} Physical
  Review B \textbf{81}, 081305 (2010).

\bibitem{Shinokita2010}
K.~Shinokita, H.~Hirori, M.~Nagai, N.~Satoh, Y.~Kadoya, and K.~Tanaka,
  \enquote{{Dynamical Franz-Keldysh effect in GaAs/AlGaAs multiple quantum
  wells induced by single-cycle terahertz pulses},} Applied Physics Letters
  \textbf{97}, 211902 (2010).

\bibitem{Hoffmann2009a}
M.~C. Hoffmann, N.~C. Brandt, H.~Y. Hwang, K.-L. Yeh, and K.~A. Nelson,
  \enquote{{Terahertz Kerr effect},} Applied Physics Letters \textbf{95},
  231105 (2009).

\bibitem{Fleischer2011}
S.~Fleischer, Y.~Zhou, R.~W. Field, and K.~A. Nelson, \enquote{{Molecular
  Orientation and Alignment by Intense Single-Cycle THz Pulses},} Physical
  Review Letters \textbf{107}, 163603 (2011).

\bibitem{Fleischer2012}
S.~Fleischer, R.~W. Field, and K.~A. Nelson, \enquote{{Commensurate Two-Quantum
  Coherences Induced by Time-Delayed THz Fields},} Physical Review Letters
  \textbf{109}, 123603 (2012).

\bibitem{Jewariya2010}
M.~Jewariya, M.~Nagai, and K.~Tanaka, \enquote{{Ladder Climbing on the
  Anharmonic Intermolecular Potential in an Amino Acid Microcrystal via an
  Intense Monocycle Terahertz Pulse},} Physical Review Letters \textbf{105},
  203003 (2010).

\bibitem{Grady2013}
N.~K. Grady, B.~G. Perkins, H.~Y. Hwang, N.~C. Brandt, D.~Torchinsky, R.~Singh,
  L.~Yan, D.~Trugman, S.~A. Trugman, Q.~X. Jia, A.~J. Taylor, K.~A. Nelson, and
  H.-T. Chen, \enquote{{Nonlinear high-temperature superconducting terahertz
  metamaterials},} New Journal of Physics \textbf{15}, 105016 (2013).

\bibitem{Liu2012}
M.~Liu, H.~Y. Hwang, H.~Tao, A.~C. Strikwerda, K.~Fan, G.~R. Keiser, A.~J.
  Sternbach, K.~G. West, S.~Kittiwatanakul, J.~Lu, S.~A. Wolf, F.~G. Omenetto,
  X.~Zhang, K.~A. Nelson, and R.~D. Averitt, \enquote{{Terahertz-field-induced
  insulator-to-metal transition in vanadium dioxide metamaterial.}} Nature
  \textbf{487}, 345--348 (2012).

\bibitem{vanderMeer2004}
A.~van~der Meer, \enquote{{FELs, nice toys or efficient tools?}} Nuclear
  Instruments and Methods in Physics Research Section A: Accelerators,
  Spectrometers, Detectors and Associated Equipment \textbf{528}, 8--14 (2004).

\bibitem{Williams2007}
B.~S. Williams, \enquote{{Terahertz quantum-cascade lasers},} Nature Photonics
  \textbf{1}, 517--525 (2007).

\bibitem{Hebling2008}
J.~Hebling, K.-L. Yeh, M.~C. Hoffmann, B.~Bartal, and K.~A. Nelson,
  \enquote{{Generation of high-power terahertz pulses by tilted-pulse-front
  excitation and their application possibilities},} Journal of the Optical
  Society of America B \textbf{25}, B6 (2008).

\bibitem{Auston1984}
D.~H. Auston, K.~Cheung, J.~Valdmanis, and D.~A. Kleinman, \enquote{{Cherenkov
  Radiation from Femtosecond Optical Pulses in Electro-Optic Media},} Physical
  Review Letters \textbf{53}, 1555--1558 (1984).

\bibitem{Hebling2002}
J.~Hebling, G.~Almasi, I.~Kozma, and J.~Kuhl, \enquote{{Velocity matching by
  pulse front tilting for large area THz-pulse generation.}} Optics Express
  \textbf{10}, 1161--1166 (2002).

\bibitem{Stepanov2003}
A.~G. Stepanov, J.~Hebling, and J.~Kuhl, \enquote{{Efficient generation of
  subpicosecond terahertz radiation by phase-matched optical rectification
  using ultrashort laser pulses with tilted pulse fronts},} Applied Physics
  Letters \textbf{83}, 3000 (2003).

\bibitem{Hebling1996}
J.~Hebling, \enquote{{Derivation of the pulse front tilt caused by angular
  dispersion},} Optical and Quantum Electronics \textbf{28}, 1759--1763 (1996).

\bibitem{Kunitski2013}
M.~Kunitski, M.~Richter, M.~D. Thomson, A.~Vredenborg, J.~Wu, T.~Jahnke,
  M.~Sch\"{o}ffler, H.~Schmidt-B\"{o}cking, H.~G. Roskos, and R.~D\"{o}rner,
  \enquote{{Optimization of single-cycle terahertz generation in LiNbO$_{3}$
  for sub-50 femtosecond pump pulses},} Optics Express \textbf{21}, 6826--6836
  (2013).

\bibitem{Yeh2007}
K.-L. Yeh, M.~C. Hoffmann, J.~Hebling, and K.~A. Nelson, \enquote{{Generation
  of 10 $\mu$J ultrashort terahertz pulses by optical rectification},} Applied
  Physics Letters \textbf{90}, 171121 (2007).

\bibitem{Stepanov2008}
A.~G. Stepanov, L.~Bonacina, S.~V. Chekalin, and J.-p. Wolf,
  \enquote{{Generation of 30 $\mu$J single-cycle terahertz pulses at 100 Hz
  repetition rate by optical rectification},} Optics Letters \textbf{33}, 2497
  (2008).

\bibitem{Nagai2009}
M.~Nagai, M.~Jewariya, Y.~Ichikawa, H.~Ohtake, T.~Sugiura, Y.~Uehara, and
  K.~Tanaka, \enquote{{Broadband and high power terahertz pulse generation
  beyond excitation bandwidth limitation via $\chi^{2}$ cascaded processes in
  LiNbO$_{3}$},} Optics Express \textbf{17}, 11543 (2009).

\bibitem{Bodrov2013}
S.~B. Bodrov, A.~A. Murzanev, Y.~A. Sergeev, Y.~A. Malkov, and A.~N. Stepanov,
  \enquote{{Terahertz generation by tilted-front laser pulses in weakly and
  strongly nonlinear regimes},} Applied Physics Letters \textbf{103}, 251103
  (2013).

\bibitem{Huang2013}
S.-W. Huang, E.~Granados, W.~R. Huang, K.-H. Hong, L.~E. Zapata, and F.~X.
  K\"{a}rtner, \enquote{{High conversion efficiency, high energy terahertz
  pulses by optical rectification in cryogenically cooled lithium niobate.}}
  Optics Letters \textbf{38}, 796--8 (2013).

\bibitem{Fulop2014}
J.~A. F\"{u}l\"{o}p, Z.~Ollmann, C.~Lombosi, C.~Skrobol, S.~Klingebiel,
  L.~P\'{a}lfalvi, F.~Krausz, S.~Karsch, and J.~Hebling, \enquote{{Efficient
  generation of THz pulses with 0.4 mJ energy},} Optics Express \textbf{22},
  20155--20163 (2014).

\bibitem{Huang2014}
W.~R. Huang, S.-W. Huang, E.~Granados, K.~Ravi, K.-H. Hong, L.~E. Zapata, and
  F.~X. K\"{a}rtner, \enquote{{Highly efficient terahertz pulse generation by
  optical rectification in stoichiometric and cryo-cooled congruent lithium
  niobate},} Journal of Modern Optics \textbf{62}, 1486 (2014).

\bibitem{Wu2014}
X.~Wu, S.~Carbajo, K.~Ravi, F.~Ahr, G.~Cirmi, Y.~Zhou, O.~D. M\"{u}cke, and
  F.~X. K\"{a}rtner, \enquote{{Terahertz generation in lithium niobate driven
  by Ti:sapphire laser pulses and its limitations},} Optics Letters
  \textbf{39}, 5403 (2014).

\bibitem{Fulop2010}
J.~A. F\"{u}l\"{o}p, L.~P\'{a}lfalvi, G.~Almasi, and J.~Hebling,
  \enquote{{Design of high-energy terahertz sources based on optical
  rectification.}} Optics Express \textbf{18}, 12311--27 (2010).

\bibitem{Fulop2011}
J.~A. F\"{u}l\"{o}p, L.~P\'{a}lfalvi, M.~C. Hoffmann, and J.~Hebling,
  \enquote{{Towards generation of mJ-level ultrashort THz pulses by optical
  rectification},} Optics Express \textbf{19}, 15090--15097 (2011).

\bibitem{Ravi2014}
K.~Ravi, W.~R. Huang, S.~Carbajo, X.~Wu, and F.~K\"{a}rtner,
  \enquote{{Limitations to THz generation by optical rectification using tilted
  pulse fronts},} Optics Express \textbf{22}, 20239 (2014).

\bibitem{Martinez1986}
O.~E. Martinez, \enquote{{Pulse distortions in tilted pulse schemes for
  ultrashort pulses},} Optics Communications \textbf{59}, 229--232 (1986).

\bibitem{Lombosi2015}
C.~Lombosi, G.~Pol\'{o}nyi, M.~Mechler, Z.~Ollmann, J.~Hebling, and J.~A.
  F\"{u}l\"{o}p, \enquote{{Nonlinear distortion of intense THz beams},} New
  Journal of Physics \textbf{17}, 083041 (2015).

\bibitem{Palfalvi2008}
L.~P\'{a}lfalvi, J.~A. F\"{u}l\"{o}p, G.~Almasi, and J.~Hebling, \enquote{{Novel
  setups for extremely high power single-cycle terahertz pulse generation by
  optical rectification},} Applied Physics Letters \textbf{92}, 171107 (2008).

\bibitem{Diels2006}
J.-C. Diels and W.~Rudolph, \emph{{Ultrashort Laser Pulse Phenomena}} (Academic
  Press, San Diego, CA, 2006), 2nd ed.

\bibitem{Minami2013}
Y.~Minami, Y.~Hayashi, J.~Takeda, and I.~Katayama, \enquote{{Single-shot
  measurement of a terahertz electric-field waveform using a reflective echelon
  mirror},} Applied Physics Letters \textbf{103}, 051103 (2013).

\bibitem{Stepanov2005a}
A.~G. Stepanov, J.~Hebling, and J.~Kuhl, \enquote{{THz generation via optical
  rectification with ultrashort laser pulse focused to a line},} Applied
  Physics B \textbf{81}, 23--26 (2005).

\bibitem{Feurer2007}
T.~Feurer, N.~S. Stoyanov, D.~W. Ward, J.~C. Vaughan, E.~R. Statz, and K.~A.
  Nelson, \enquote{{Terahertz Polaritonics},} Annual Review of Materials
  Research \textbf{37}, 317--350 (2007).

\bibitem{Blanchard2014}
F.~Blanchard, X.~Ropagnol, H.~Hafez, H.~Razavipour, M.~Bolduc, R.~Morandotti,
  T.~Ozaki, and D.~G. Cooke, \enquote{{Effect of extreme pump pulse reshaping
  on intense terahertz emission in lithium niobate at multimilliJoule pump
  energies},} Optics Letters \textbf{39}, 4333--4336 (2014).

\bibitem{Back1954}
F.~G. Back and H.~Lowen, \enquote{{The Basic Theory of Varifocal Lenses with
  Linear Movement and Optical Compensation},} Journal of the Optical Society of
  America \textbf{44}, 684--691 (1954).

\bibitem{Koehl2001}
R.~M. Koehl and K.~A. Nelson, \enquote{{Coherent optical control over
  collective vibrations traveling at lightlike speeds},} The Journal of
  Chemical Physics \textbf{114}, 1443--1446 (2001).

\bibitem{Koehl2001a}
R.~M. Koehl and K.~A. Nelson, \enquote{{Terahertz polaritonics: Automated
  spatiotemporal control over propagating lattice waves},} Chemical Physics
  \textbf{267}, 151--159 (2001).

\bibitem{Feurer2003}
T.~Feurer, J.~C. Vaughan, and K.~A. Nelson, \enquote{{Spatiotemporal coherent
  control of lattice vibrational waves.}} Science \textbf{299}, 374--377
  (2003).

\bibitem{Yeh2007a}
K.-L. Yeh, T.~Hornung, J.~C. Vaughan, and K.~A. Nelson, \enquote{{Terahertz
  amplification in high-dielectric materials},} in \enquote{Ultrafast Phenomena
  XV Springer Series in Chemical Physics,} , vol.~88, P.~Corkum, D.~Jonas,
  R.~Miller, and A.~M. Weiner, eds. (2007), vol.~88, pp. 802--804.

\bibitem{Werley2011}
C.~A. Werley, S.~M. Teo, and K.~A. Nelson, \enquote{{Pulsed laser noise
  analysis and pump-probe signal detection with a data acquisition card.}}
  Review of Scientific Instruments \textbf{82}, 123108 (2011).

\bibitem{Hwang2014}
H.~Y. Hwang, S.~Fleischer, N.~C. Brandt, B.~G. Perkins, M.~Liu, K.~Fan,
  A.~Sternbach, X.~Zhang, R.~D. Averitt, and K.~A. Nelson, \enquote{{A review
  of non-linear terahertz spectroscopy with ultrashort tabletop-laser pulses},}
  Journal of Modern Optics \textbf{62}, 1447 (2014).

\bibitem{Hebling2004}
J.~Hebling, A.~G. Stepanov, G.~Almasi, B.~Bartal, and J.~Kuhl,
  \enquote{{Tunable THz pulse generation by optical rectification of ultrashort
  laser pulses with tilted pulse fronts},} Applied Physics B \textbf{78},
  593--599 (2004).
  
\bibitem{Reid2005}
M.~Reid and R.~Fedosejevs, \enquote{{Quantitative comparison of terahertz
  emission from (100) InAs surfaces and a GaAs large-aperture photoconductive
  switch at high fluences.}} Applied optics \textbf{44}, 149--153 (2005).

\bibitem{Nagai2012}
M.~Nagai, E.~Matsubara, and M.~Ashida, \enquote{{High-efficiency terahertz
  pulse generation via optical rectification by suppressing stimulated Raman
  scattering process},} Optics Express \textbf{20}, 6509--6514 (2012).

\bibitem{Stepanov2007}
A.~G. Stepanov, A.~A. Mel'nikov, V.~O. Kompanets, and S.~V. Chekalin,
  \enquote{{Spectral modification of femtosecond laser pulses in the process of
  highly efficient generation of terahertz radiation via optical
  rectification},} JETP Letters \textbf{85}, 227--230 (2007).

\bibitem{Jewariya2009}
M.~Jewariya, M.~Nagai, and K.~Tanaka, \enquote{{Enhancement of terahertz wave
  generation by cascaded $\chi^{2}$ processes in LiNbO$_{3}$},} Journal of
  the Optical Society of America B \textbf{26}, A101 (2009).

\bibitem{Cronin-Golomb2004}
M.~Cronin-Golomb, \enquote{{Cascaded nonlinear difference-frequency generation
  of enhanced terahertz wave production},} Optics Letters \textbf{29}, 2046
  (2004).

\bibitem{Hansen2009}
R.~C. Hansen, \emph{{Phased Array Antennas}} (John Wiley \& Sons, Hoboken, N.J.,
  2009), 2nd ed.

\bibitem{Hoffmann2007}
M.~C. Hoffmann, K.-L. Yeh, J.~Hebling, and K.~A. Nelson, \enquote{{Efficient
  terahertz generation by optical rectification at 1035 nm},} Optics Express
  \textbf{15}, 11706 (2007).

\end{thebibliography}

\end{document}